\documentclass[10pt, twocolumn]{svjour3}

\usepackage{url}
\usepackage{times}
\usepackage{latexsym}
\usepackage{xspace}
\usepackage{amsmath,amssymb,amsfonts}
\usepackage{stmaryrd}
\usepackage{epstopdf}
\usepackage{eurosym}

{\qed \end{trivlist}}
\renewcommand{\qed}{\hspace*{\fill} \rule{1.1mm}{2.2mm}}

{\qed \end{trivlist}}

\newenvironment{theorem*}[2]%
{\begin{trivlist} \item[] {\bf #1~\protect{\ref{#2}.}}\it}{\end{trivlist}}

\newcommand{\hide}[1]{}

\newcommand{\easytt}[1]{\ensuremath{\mathtt{#1}}\xspace}

\newcommand{\tab}[1]{\hspace*{#1em}}

\newcommand{\PL}{\ensuremath{\mathcal{PL}}\xspace}

\usepackage{caption}
\usepackage{graphicx}
\usepackage{paralist}
\usepackage{xcolor}
\usepackage{ifthen}
\usepackage[numbers,sort]{natbib}
\usepackage{setspace}
\usepackage{dirtytalk}

\newboolean{showcomments}
\setboolean{showcomments}{true} 
\ifthenelse{\boolean{showcomments}}
  {\newcommand{\nb}[2]{
    \fcolorbox{gray}{yellow}{\bfseries\sffamily\scriptsize#1}
    {$\blacktriangleright$#2$\blacktriangleleft$}
   }
   
  }
  {\newcommand{\nb}[2]{}
   
  }

\usepackage{url}

\begin{document}
\title{Machine Understandable Policies and GDPR Compliance Checking}

\author{Piero A.\ Bonatti \and Sabrina Kirrane \and Iliana M.\ Petrova \and Luigi Sauro}


\institute{Piero Bonatti \and Iliana M.\ Petrova \and Luigi Sauro \at Universit\`a di Napoli Federico II, Naples, Italy\\
              \email{pab@unina.it}         
           \and
           Sabrina Kirrane \at
              Vienna University of Economics and Business, Vienna, Austria \\
              \email{sabrina.kirrane@wu.ac.at}                                      
}

\date{Received: 31st October 2019 / Accepted: tba}

\maketitle

\begin{abstract}
The European General Data Protection Regulation (GDPR) calls for technical and organizational measures to support its implementation.
Towards this end, the SPECIAL H2020 project aims to provide a set of tools that can be used by data controllers and processors to automatically check if personal data processing and sharing complies with the obligations set forth in the GDPR.
The primary contributions of the project include: (i) a policy language that can be used to express consent, business policies, and regulatory obligations; and (ii) two different approaches to automated compliance checking that can be used to demonstrate that data processing performed by data controllers / processors complies with consent provided by data subjects, and business processes comply with regulatory obligations set forth in the GDPR. 
\end{abstract}

%
\maketitle

\section{Introduction}
\label{sec:introduction}
The European General Data Protection
Regulation (GDPR)
, which came into force on the 25th of May 2018, defines legal requirements concerning the processing and sharing of personally identifiable data.
In addition, the legislation calls for technical and organizational measures to support its implementation.

When it comes to legal informatics there is a large body of work on legal knowledge representation and reasoning (cf., \cite{bartolini2015using,pandit2018gdprtext, palmirani2011legalruleml, athan2013oasis, governatori2016semantic, lam2019enabling}), however said approaches are usually foundational in nature and as such are not readily accessible for companies looking for technical means to demonstrate GDPR compliance. 

Recently we have seen the emergence of GDPR compliance tools (cf., \cite{ICO2017,MicrosoftTrustCenter2017,Nymity,agarwal2018legislative}) in the form of predefined questionnaires that enable data controllers and processors to assess the compliance of services and products that process personal data. The primary limitation of said tools is their lack of support for automated compliance checking. 

In order to fill this gap, SPECIAL builds upon a rich history of policy language research from the Semantic Web community (cf., \cite{DBLP:journals/csec/WooL93,DBLP:journals/tods/JajodiaSSS01,DBLP:conf/policy/UszokBJSHBBJKL03,rei,DBLP:journals/tkde/BonattiCOS10}), and shows how together machine understandable policies and automated compliance checking can be used to demonstrate compliance with legal requirements set forth in the GDPR.

In particular, we introduce the SPECIAL policy language and discuss how it can be used to express consent, business policies, and regulatory obligations. In addition, we describe two different approaches to automated compliance checking used to demonstrate that: (i) data processing performed by data controllers / processors complies with consent provided by data subjects; and (ii) business processes comply with regulatory obligations set forth in the GDPR. In addition, we provide a highlevel overview of our compliance checking algorithm and present the results of our initial performance evaluation.

The remainder of the paper is structured as follows:  
\emph{Section}~\ref{sec:structural-analysis} describes our analysis of the text of the GDPR.
\emph{Section}~\ref{sec:consent} introduces the SPECIAL policy language, which provides a machine understandable encoding of consent.
\emph{Section}~\ref{sec:business-policies} discusses how the SPECIAL policy language can be used to encode business policies and regulatory obligations. 
\emph{Section}~\ref{sec:compliance} presents our compliance checking algorithm and the results of our initial performance evaluation. 
\emph{Section}~\ref{sec:related-work} points to related work on GDPR compliance.  
Finally, we present our conclusions and interesting directions for future work in \emph{Section}~\ref{sec:conclusions}.

\section{Requirements Analysis}
\label{sec:structural-analysis}

One of the primary goals of SPECIAL is to automatically check if personal data processing and sharing performed by data controllers and processors complies with obligations set forth in the GDPR.
A necessary first step is to better understand the text of the GDPR, its interpretation by legal professionals, and the role of machine understandable representations, and automated compliance checking.

\subsection{GDPR Analysis}
\label{sec:GDPR-analysis}

Legal rules are composed of several constructs, prohibitions (used to describe what is not permitted), permissions (used to describe what is permitted), obligations (used to describe requirements that must be fulfilled), and dispensations (used to describe exemptions), commonly referred to as deontic concepts. In addition to these common constructs, the legal language contains constraints (used to limit the scope of permissions, prohibitions, obligations and dispensations), definitions (used to establish meaning), dispositions (used to highlight best practices/ suggestions), and opening clauses (used to indicate the need to consult National or European legislation). 
%
%
When it comes to encoding legislative requirements using machine understandable representations, such that it is possible to perform automated compliance, major considerations include:

\paragraph{Connectedness} of the various articles, paragraphs, and points, which can either explicitly refer to another piece of legislation (e.g., \emph{\say{scientific or historical research purposes or statistical purposes in accordance with Article 89(1)}}) or implicitly to knowledge about the law (e.g., \emph{\say{Personal data shall be: (a) processed lawfully, fairly and in a transparent manner in relation to the data subject (’lawfulness, fairness and transparency’)}}).
In either case, from an automated compliance checking perspective, it is clear that legal requirements are not separate and distinct rules but rather rules need to be linked, clustered, and/or generalized in a manner that enables the validation of a combination rules.
In SPECIAL we do not try to encode the entire GDPR, but rather focus on encoding legislative obligations (relating to several articles, paragraphs, and points) such that: (i) data processing performed by data controllers / processors complies with consent provided by data subjects; and (ii) business processes comply with regulatory obligations set forth in the GDPR.   

\paragraph{Temporal expressions} provide contextual information that is relevant for the interpretation of actions that need to be taken. Several different types of temporal expressions can be found in the text of the GDPR, for instance:
\begin{itemize}
\item \dots \emph{\say{the right to withdraw his or her consent at any time}} (Article 7 paragraph 3);    
\item \dots \emph{\say{processing based on consent before its withdrawal}} (Article 7 paragraph 3, Article 13 paragraph 2, Article 14 paragraph 2);
\item \dots \emph{\say{prior to giving consent}} (Article 7 paragraph 3);   
\item \dots \emph{\say{at the time when personal data are obtained}} (Article 13 paragraphs 1 and 2);
\item \dots \emph{\say{the personal data shall no longer be processed}} (Article 21 paragraph 3); 
\end{itemize}
\noindent In SPECIAL we provide support for such temporal requirements by recording in a suitable \emph{transparency ledger} when consent was obtained or when the data processing/sharing happened.  This information is used for both ex-ante and ex-post compliance checking (as well as  other purposes, discussed later).

\subsection{Legal Interpretations}

The GDPR defines several potential legal bases (consent, contract, legal obligation, vital interest, public interest, exercise of official authority, and  legitimate interest) under which companies can legally process personal data. 
In order to determine if personal data processing is legally valid, the legal inquiry process usually involves gathering specific information such as: 
\begin{inparaenum} [(i)]
\item the \emph{personal data} collected from the data subject;
\item the \emph{processing} that are performed on the personal data;
\item the \emph{purpose} of such processing;
\item where data are \emph{stored and for how long}; and
\item with whom data is \emph{shared}.
\end{inparaenum}
The answers provided to said questions enable legal professionals to determine which articles need to be consulted in order both to assess the lawfulness of processing and to identify relevant legal obligations.

Although, the open textured nature of legal texts is a highly desirable feature, as it leaves room for interpretation on a case by case basis, such ambiguity poses challenges for automatic compliance checking. In terms of legal interpretations, legal professionals also need to interpret the facts of the case with respect to relevant National or European legislation (e.g., opening clauses) and subjective terms (e.g., single words or parts of a sentence that can be interpreted in various ways). Here legal knowledge graphs could potentially play a crucial role as they allow for the modeling of both legislation and cases in a machine readable format, based on standardization activities such as European Law Identifier (ELI) and the European Case Law Identifier (ECLI), which provide technical specifications for web identifiers and vocabularies that can be used to describe metadata pertaining to legal documents. Such a legal knowledge graph could be used not only to identify case specific legislation, but also to uncover if there have been any prior cases that could be used to reduce ambiguity.
The SPECIAL poly language has been developed together with legal professionals who well versed in the interpretation of legal texts. Going forward we envisage that legal knowledge graphs could be used to reduce subjectivity thus allowing us to perform automated compliance checking for a broader set of legislative requirements.

\subsection{Machine Understandable Representations}

The GDPR poses at least two requirements that call for a
machine-understandable representation of data usage modalities.
Article~30 states that each controller shall maintain a record of the personal data
processing activities under its responsibility.  The first paragraph
specifies that such a ledger should describe (among other information)
the following aspects of \emph{data usage}:
\begin{itemize}
\item[P1.] the \emph{purpose} of processing;
\item[P2.] a description of the \emph{categories of data subjects} and of the
  \emph{categories of personal data};
\item[P3.] the categories of \emph{recipients} to whom the personal data have
  been or will be disclosed;
\item[P4.] \emph{transfers} of personal data to a third country or an
  international organization (since cross-border data transfer are subject to limitations);
\item[P5.] the envisaged \emph{time limits for erasure} of the different
  categories of data;
\item[P6.] information about the \emph{processing}, such as the security measures
  mentioned in Article~32.
\end{itemize}

\noindent
Recital~42 stresses that, where processing is based on the data
subject's consent, the controller should be able to demonstrate that
the data subject has given consent to the processing
operation. SPECIAL addresses this issue by recording consent in the
{transparency ledger} (cf.\ Sec.~\ref{sec:GDPR-analysis}). The description of consent is similar to
the description of processing activities as per Article~30.  While
Article~6.1.(a) -- that introduces consent as a legal basis for
personal data processing -- and Recital~42 explicitly mention only the
purpose of processing, Articles~13 and 14 add the other elements
P2--P6 listed above.  Concerning P6 (processing), it should be
specified whether any automated decision making is involved, including
profiling.


\subsection{Automated Compliance Checking}

Once such data usage descriptions are encoded in a
machine-understandable way, several tasks, related to GDPR
compliance, can be automated, including:
\begin{itemize}
\item[T1.] Checking whether the processing complies with several
  restrictions imposed by the GDPR, such as additional requirements on
  the processing of sensitive data, restrictions on cross-border
  transfers, and compatibility of data usage with the chosen legal
  basis. This kind of validation requires a machine-understandable
  formalization of the relevant parts of the GDPR.

\item[T2.] Checking whether a specific operation is permitted by the
  available consent.

\item[T3.] Running ex-post auditing on the controller's activities. In
  SPECIAL this task is supported by logging data processing
  events in the transparency ledger, and comparing such events with consent.

\item[T4.] Finding the consent that justifies a specific processing (for
  auditing or responding to a data subject's inquiry).
\end{itemize}

The transparency ledger is also used in SPECIAL to provide dashboards to data
subjects, that support them in monitoring the use of their data and
\emph{explaining} why their consent allowed specific operations.  Such
dashboards can also be used as a uniform interface to let data
subjects exercise their rights (access to data, right to erasure,
etc.) as specified by Articles~15--18 and 21--22.

\section{Consent Compliance Checking}
\label{sec:consent}

Although there are several potential legal bases that could be used to lawfully process personal data, in SPECIAL we have a particular focus on consent. Thus in this section we present the SPECIAL policy language and demonstrate how it can be used to encode consent in a manner than enables automated compliance checking.

\subsection{Encoding Usage Descriptions and Consent}
\label{sec:logic}
The common structure of the activity records and of the consent forms,
consisting of properties P1--P6, is called \emph{simple (usage)
  policy} in SPECIAL. In general, both the controller's activities and
the consent of data subjects can be described by a \emph{set} of
simple usage policies (covering different data categories and purposes), called \emph{full (usage) policies}.  Each
simple policy can be specified simply by attaching to each property
P$_i$ (such as purpose, data category, recipients, etc.) a term
selected from a suitable \emph{vocabulary} (ontology).

\begin{example}
    \label{ex:befit}
  A company -- call it BeFit -- sells a wearable fitness
  appliance and wants (i) to process biometric data (stored in the EU) for
  sending health-related advice to its customers, and (ii) share the customer's location
  data with their friends. Location data are kept for a minimum of one year but no longer
  than 5; biometric data are kept for an unspecified amount of time. In order to do all
  this legally, BeFit needs consent from its customers.
  Consent can be represented with two simple policies, specified using SPECIAL's vocabularies:
  {\small%
\begin{verbatim}
  {
    has_purpose: FitnessRecommendation,
    has_data: BiometricData,
    has_processing: Analytics,
    has_recipient: BeFit,
    has_storage: { has_location: EU }
  }
\end{verbatim}
\begin{verbatim}
  {
    has_purpose: SocialNetworking,
    has_data: LocationData,
    has_processing: Transfer,
    has_recipient: DataSubjFriends,
    has_storage: {
        has_location: EU,
        has_duration: [1year,5year]
        }
  }
\end{verbatim}
 }
\medskip\noindent If \texttt{HeartRate} is a
subclass of \texttt{BiometricData} and \texttt{ComputeAvg} is
a subclass of \texttt{Analytics}, then the above consent allows
BeFit to compute the average heart rate of the data subject in order
to send her fitness recommendations.
BeFit customers may restrict their consent, e.g.\ by picking a
specific recommendation modality, like ``recommendation via
SMS only''. Then the first line should be replaced with something like:
{\small%
\begin{verbatim}
  has_purpose:{
    FitnessRecommendation, 
    contact: SMS}
\end{verbatim}
}
\medskip\noindent 
\linespread{.8}\selectfont{
Moreover, a customer of BeFit may consent to the first or the second argument of
the union, or both.  Their consent would be encoded, respectively,
with the first simple policy, the second simple policy, or both.  Similarly, each single process in the controller's business application may use only biometric data, only location data, or
both.  Accordingly, it may be associated to the first simple policy,
the second simple policy, or both. 
}
\qed \end{example}


The temporary exemplifying policy language vocabularies reported in SPECIAL's deliverables have been obtained by adapting previous
standardized terms introduced by initiatives related to privacy and digital rights management, such as
P3P\footnote{\url{http://www.w3.org/TR/P3P11}} and
ODRL,\footnote{\url{https://www.w3.org/TR/odrl/}}. More refined
vocabularies have been recently proposed by W3C's \emph{Data Privacy
  Vocabularies and Controls Community Group},
(DPVCG)~\cite{Pandit2019OTM}, promoted by SPECIAL and spanning a range of stakeholders wider than the project's consortium.
The current vocabularies can be found on DPVCG's website\footnote{\url{www.w3.org/community/dpvcg/}}.

As shown in Example~\ref{ex:befit}, usage policies can be formatted
with a minor extension of JSON (in particular, compound terms and
policy sets require additional operators), while  vocabularies can be
encoded in RDFS or lightweight profiles of OWL2 such as OWL2-EL and
OWL2-QL.

A grammar for SPECIAL policy expressions in Backus–Naur form (BNF) format is presented in Figure~\ref{upl-BNF}.  The categories \textsf{\small DataVocabExpression, PurposeVocabExpression, ProcessingVocabExpression, RecipientVocabExpression, LocationVocabExpression, DurationVocabExpression} are specified by DPVCG's vocabularies.

\begin{figure}
  \caption{SPECIAL's Usage Policy Language Grammar}
  \label{upl-BNF}
  \newcommand{\is}{\ensuremath{\mathbf{\mathrel{\mathop :}=}}\xspace}
  
  \textsf{\footnotesize \small 
    \textbf{UsagePolicy} \is  'ObjectUnionOf' '('
    \textbf{BasicUsagePolicy} \\ \{
    \textbf{BasicUsagePolicy} \}* ')' \\
    \hspace*{1em} \textbf{$\;\mid\;$ BasicUsagePolicy}
    \\[\medskipamount]
    \textbf{BasicUsagePolicy} \is 'ObjectIntersectionOf' '('
    \textbf{Data Purpose Processing Recipients Storage} ')'
    \\[\smallskipamount]
    \textbf{Data} \is 'ObjectSomeValueFrom' '(' 'spl:hasData' \textbf{DataExpression} ')'
    \\[\smallskipamount]
    \textbf{Purpose} \is 'ObjectSomeValueFrom' '(' 'spl:hasPurpose' \textbf{PurposeExpression} ')'
    \\[\smallskipamount]
    \textbf{Processing} \is  'ObjectSomeValueFrom' '(' 'spl:hasProcessing' \textbf{ProcessingExpression} ')'
    \\[\smallskipamount]
    \textbf{Recipients} \is  'ObjectSomeValueFrom' '(' 'spl:hasRecipient'  \textbf{RecipientExpression} ')'
    \\[\smallskipamount]
    \textbf{Storage} \is  'ObjectSomeValueFrom' '(' 'spl:hasStorage'  \textbf{StorageExpression} ')'
    \\[\smallskipamount]
    \textbf{DataExpression} \is 'spl:AnyData' \textbf{$\;\mid\;$ DataVocabExpression}
    \\[\smallskipamount]
    \textbf{PurposeExpression} \is 'spl:AnyPurpose' \textbf{$\;\mid\;$ PurposeVocabExpression}
    \\[\smallskipamount]
    \textbf{ProcessingExpression} \is 'spl:AnyProcessing' \textbf{$\;\mid\;$ ProcessingVocabExpression}
    \\[\smallskipamount]
    \textbf{RecipientsExpression} \is 'spl:AnyRecipient' \textbf{$\;\mid\;$} 'spl:Null' \textbf{$\;\mid\;$ RecipientVocabExpression}
    \\[\smallskipamount]
    \textbf{StorageExpression} \is 'spl:AnyStorage' \textbf{$\;\mid\;$} 'spl:Null' \textbf{$\;\mid\;$} \\
    \hspace*{2em} 'ObjectIntersectionOf' '(' \textbf{Location Duration} ')'
    \\[\smallskipamount]
    \textbf{Location} \is  'ObjectSomeValueFrom' '(' 'spl:hasLocation'  \textbf{LocationExpression} ')'
    \\[\smallskipamount]
    \textbf{Duration} \is 'ObjectSomeValueFrom' '(' 'spl:hasDuration'  \textbf{DurationExpression} ')' \\
    \hspace*{5em}\textbf{$\;\mid\;$} 'DataSomeValueFrom' '(' 'spl:durationInDays'  \textbf{IntervalExpression} ')'
    \\[\smallskipamount]
    \textbf{LocationExpression} \is 'spl:AnyLocation' \textbf{$\;\mid\;$ LocationVocabExpression}
    \\[\smallskipamount]
    \textbf{DurationExpression} \is 'spl:AnyDuration' \textbf{$\;\mid\;$ DurationVocabExpression}
    \\[\smallskipamount]
    \textbf{IntervalExpression} \is 'DatatypeRestriction' '(' 'xsd:integer' \textbf{LowerBound UpperBound} ')'
    \\[\medskipamount]
    \textbf{LowerBound} \is 'xsd:minInclusive' \textbf{IntegerLiteral} 
    \\[\smallskipamount]
    \textbf{UpperBound} \is 'xsd:maxInclusive' \textbf{IntegerLiteral}
    \\[\smallskipamount]
    \textbf{IntegerLiteral} \is  \textbf{stringOfDigits} '\mbox{\textasciicircum\textasciicircum}' 'xsd:integer'
    \\[\smallskipamount]
    \textbf{stringOfDigits} \is  \textit{a sequence of digits enclosed in a pair of  " (U+22)}
  }  

\end{figure}

\subsection{Compliance Checking}
\label{sec:logic-compliance}

Internally, SPECIAL's components encode also policies and the entries of
the transparency ledger with a fragment (profile) of OWL2 called
\PL (policy logic) \cite{DBLP:conf/ijcai/Bonatti18}. The adoption of a
logic-based description language has manifold reasons.  First, it has
a clean, unambiguous semantics, that is a must for policy languages.
A formal approach brings the following advantages:
\begin{itemize}
\item strong correctness and completeness guarantees on the algorithms
  for permission checking and compliance checking;
\item the mutual coherence of the different reasoning tasks related to
  policies, such as policy validation, permission checking, compliance
  checking, and explanations (cf.\ tasks T1--T4 and the subsequent
  paragraph);
\item correct usage after data is transferred to other controllers (i.e. interoperability). When it comes to so-called \emph{sticky policies} \cite{DBLP:journals/computer/PearsonM11}, that
constitute a sort of a license that applies to the data released to
third parties, it is essential that all parties understand the sticky
policy in the same way.

\end{itemize}

Policies are modeled as OWL2 \emph{classes}. If the policy describes
a controller's activity, then its instances represent all the
operations that the controller may possibly execute.  If the policy
describes consent, then its instances represent all the operations
permitted by the data subject.  A description of (part of) the
controller's activity -- called \emph{business policy} in SPECIAL (possibly represented as a \emph{transparency log entry}) --
\emph{complies} with a consent policy if the former is a subclass of
the latter, that is, all the possible operations described by the
business policies are also permitted by the given consent.

  

  


\begin{example}
  Consider again Example~\ref{ex:befit}. The JSON-like representation
  used there can be directly mapped onto an OWL2 class
  \verb|ObjectUnionOf(|$P_1\ P_2$\verb|)|, where $P_2$ is\footnote{We omit $P_1$ due to space limitations; the reader may easily derive it by analogy with the above example.}:
  
{\medskip\small%
\begin{verbatim}
ObjectIntersectionOf(
  ObjectSomeValueFrom(
    has_purpose SocialNetworking )
  ObjectSomeValueFrom(
    has_data LocationData)
  ObjectSomeValueFrom(
    has_processing Transfer)
  ObjectSomeValueFrom(
    has_recipient DataSubjFriends)
  ObjectSomeValueFrom(
    has_storage ObjectIntersectionOf(
      ObjectSomeValueFrom(has_location: EU)
      DataSomeValueFrom(has_duration
        DatatypeRestriction(xsd:integer 
          xsd:minInclusive "365"^^xsd:integer
          xsd:maxInclusive "1825"^^xsd:integer
)))
\end{verbatim}}

\medskip\noindent

In order to check whether a business policy $BP$ (encoded as an OWL2
class) complies with the above policy one should check whether the former is a subclass of the latter, that is, whether:
\begin{center}
{\small
  \verb|SubClassOf(|$BP$ \verb|ObjectUnionOf(|$P_1\ P_2$\verb|))|
}
\end{center}
is a logical consequence of the ontology that defines SPECIAL's
vocabularies.  \qed
\end{example}

\section{Business Processes Compliance Checking}
\label{sec:business-policies}
\begin{figure}
  \caption{SPECIAL's Business Policy Language Grammar}
  \label{BP-BNF}
  \newcommand{\is}{\ensuremath{\mathbf{\mathrel{\mathop :}=}}\xspace}

  \textsf{\footnotesize \small 
    \textbf{BusinessPolicy} \is  \textbf{BasicBP $\;\mid\;$}\\ \hspace*{2em} 'ObjectUnionOf' '('
    \textbf{BasicBP} \{
    \textbf{BasicBP} \}* ')' 
    \\[\smallskipamount]
    \textbf{BasicBP} \is 'ObjectIntersectionOf' '('
    \textbf{Data Purpose Processing Recipients Storage \{Duty\}* \{LegalBasis\}} ')'
    \\[\smallskipamount]
    \textbf{Data} \is \textit{see \emph{Section}~\ref{sec:consent}}
    \\[\smallskipamount]
    \textbf{Purpose} \is  \textit{see \emph{Section}~\ref{sec:consent}}
    \\[\smallskipamount]
    \textbf{Processing} \is   \textit{see \emph{Section}~\ref{sec:consent}}
    \\[\smallskipamount]
    \textbf{Recipients} \is   \textit{see \emph{Section}~\ref{sec:consent}}
    \\[\smallskipamount]
    \textbf{Storage} \is   \textit{see \emph{Section}~\ref{sec:consent}}
    \\[\smallskipamount]
    \textbf{Duty} \is   'ObjectSomeValuesFrom' '(' 'sbpl:hasDuty' \textbf{DutyExpression} ')'
    \\[\smallskipamount]
    \textbf{DutyExpression} \is   'sbpl:AnyDuty' \textbf{$\;\mid\;$ DutyVocabExpression} 
    \\[\smallskipamount]
    \textbf{LegalBasis} \is   'ObjectSomeValuesFrom' '(' 'sbpl:hasLegalBasis \textbf{LegalBasisVocabExpression} ')'
  }
\end{figure}


Beyond consent, the GDPR defines obligations that apply to the data controllers / processors internal
systems and processes. Here are two examples:

\begin{itemize}
\item whenever the data controller operates on personal data, it must
  \emph{acquire explicit consent} from the involved data subjects, unless the
  purpose of data processing falls within a set of exceptional cases
  (e.g.\ the processing is required by law); cf.\ Article~6.1, (b)--(f);

\item whenever data are transferred to a third country whose data protection regulations do not match
  the EU requirements, alternative guarantees must be provided, e.g.\
  in the form of company regulations called \emph{binding corporate
  rules}, cf.\ Article~47 and, more generally, GDPR Chapter~V (Transfer Of
  Personal Data To Third Countries Or International Organisations).
\end{itemize}
Moreover, and differently from the above examples, the GDPR sets obligations that are not directly related to the controller's business processes, such as the requirement that data subjects have the right to \emph{access, rectify, and delete} their personal data.  In order to fulfill such obligations, data controllers have to set up suitable processes.
Last but not least, it is useful to label the controller/processors processes with the legal basis for the processing; this helps in assessing and demonstrating the lawfulness of data processing activities.
For automated compliance checking descriptions of internal systems and processes should be adequately formalized in a machine-understandable way; moreover, the formalization should represent accurately the real processes, in order to make the automated compliance verification reliable.

\subsection{Encoding Business Processes as Policies}

In SPECIAL, we address a concrete setting 
in which a partial and abstract description of processes is available.  Each process description is shaped like a \emph{formalized business policy} consisting of the following set of features:
\begin{itemize}
\item the file(s) to be processed;
\item the software that carries out the processing;
\item the purpose of the processing;
\item the entities that can access the results of the processing;
\item the details of where the results are stored and for how long;
\item \emph{the obligations that are fulfilled while (or before) carrying out the processing;}
\item \emph{the legal basis of the processing.}
\end{itemize}
It is not hard to see that the first five elements in the above list
match  SPECIAL's {usage
  policy language} (UPL) introduced in \emph{Section}~\ref{sec:consent}.  As far as the above elements are concerned, the
only difference between UPL expressions and a business policy is the
granularity of attribute values.  For example, the involved data
(specified in the first element of the above list) are not expressed as a
general, content-oriented category, but rather as a concrete set of
data sources or data items.  Such objects can be modeled as instances or subclasses of the general data categories illustrated in \emph{Section}~\ref{sec:consent}, thereby creating a link between digital artifacts and usage policies.  Similar considerations hold for the other attributes:
\begin{itemize}
\item processing is not necessarily described in the abstract terms adopted by the
  processing vocabulary introduced in \emph{Section}~\ref{sec:consent}; in a business policy, this
  can be specified by naming concrete software procedures;

\item the purpose of data processing may be directly related to the
  data controller's mission and products;

\item recipients may consist of a concrete list of legal and/or physical
  persons, as opposed to general categories such as \texttt{Ours} or
  \texttt{ThirdParty};

\item storage may be specified by a list of specific data
  repositories, at the level of files and hosts.
\end{itemize}
With this level of granularity, specific authorizations can be derived from the business policy, for example:
\begin{quotation}
  \noindent
  \it The indicated software procedure can read the indicated data sources.  The results can be written in the specified repositories. The specified recipients can read the repositories...
\end{quotation}
This methodology for generating authorizations fosters a close correspondence between the business policy and the actual behavior of the data controller's systems and processes.

The attribute encoding obligations
is not part of usage policies.  It plays a dual role, representing:
\begin{itemize}
\item preconditions authorizations specified by the business policy, e.g.\ if the obligation is something like \texttt{getValidConsent} then the derived authorizations is a \emph{rule} like \emph{the specified software can read the data sources if consent has been given};

\item obligation assertions (under human responsibility) that the data controller has set up \emph{processes for fulfilling the indicated obligations} -- e.g.\ a process to obtain consent from the data subjects -- which is relevant to checking compliance with the GDPR.
\end{itemize}




\subsection{Business Policies in OWL2}
\label{sec:BP-OWL}

A basic business policy is simply a usage policy (as in \emph{Section}~\ref{sec:consent}) extended with zero or more obligations, and a legal basis, encoded with attributes \texttt{hasDuty} and  \texttt{hasLegalBasis},
%
%
%
%
%
%
%
%
%
%
%
%
%
%
%
%
%
%
for example the following policy associates the collection of personal demographic information to the obligations to get consent and let the data subject exercise her rights:
\begin{equation*}
  \begin{minipage}[t]{.9\textwidth}
    \tt\small
    ObjectIntersectionOf(
    
    ~~~ObjectSomeValuesFrom
    		
    		~~~~~~(\textbf{spl:hasData} {\tt svd:Demographic})

    ~~~ObjectSomeValuesFrom
    		
    		~~~~~~(\textbf{spl:hasProcessing} {\tt svpr:Collect})
    
    ~~~ObjectSomeValuesFrom
    		
    		~~~~~~(\textbf{spl:hasPurpose} {\tt svpu:Account})
    
    ~~~ObjectSomeValuesFrom
    		
    		~~~~~~(\textbf{spl:hasRecipient} {\tt svr:Ours})
    
    ~~~ObjectSomeValuesFrom
    		
    		~~~~~~(\textbf{spl:hasStorage}

    ~~~~~~{\tt ObjectIntersectionOf( } 

    ~~~~~~~~~\textbf{spl:hasLocation} {\tt svl:OurServers }

    ~~~~~~~~~\textbf{spl:hasDuration} {\tt svdu:Indefinitely }

    ~~~~~~{\tt  )}

    ~~~{\tt  )}
    
    ~~~ObjectSomeValuesFrom
    		
    		~~~~~~(\textbf{sbpl:hasDuty} {\it getValidConsent})
    
    ~~~ObjectSomeValuesFrom
    		
    		~~~~~~(\textbf{sbpl:hasDuty} {\it getAccessReqs})
    
    ~~~ObjectSomeValuesFrom
    		
    		~~~~~~(\textbf{sbpl:hasDuty} {\it getRectifyReqs})
    
    ~~~ObjectSomeValuesFrom
    		
    		~~~~~~(\textbf{sbpl:hasDuty} {\it getDeleteReqs})
    
    ~~~ObjectSomeValuesFrom
    		
    		~~~~~~(\textbf{sbpl:hasLegalBasis} {\it A6-1-a-explicit-consent})
    
    )
  \end{minipage}
\end{equation*}
%

Similarly to usage policies, \emph{general} business policies can be composed by enclosing several basic business policies inside the \texttt{ObjectUnionOf} operator of OWL2.
The syntax and the logical semantics of SPECIAL's Business Policy Language are specified in Figure~\ref{BP-BNF}. The values for attributes \textsf{\small DutyVocabExpression} and  \textsf{\small LegalBasisVocabExpression} are specified in DPVCG's vocabularies.

\subsection{Partial Encoding of the GDPR in OWL2}

The GDPR cannot be fully axomatized due to the usual difficulties that
arise in axiomatizing legal text (especially the frequent use of subjective
terms as highlighted in \emph{Section}~\ref{sec:structural-analysis}). However it is possible to encode some constraints that should
hold over the different attributes of a business policy.  At the top level, the formalization is organized as follows:

\begin{center}
  \begin{minipage}{30em}
    \small\tt
    ObjectUnionOf(\\
    \tab{1} ObjectIntersectionOf(\\
    \tab{2} Chap2\_LawfulProcessing\\
    \tab{2} Chap3\_RightsOfDataSubjects\\
    \tab{2} Chap4\_ControllerAndProcessorObligations\\
    \tab{2} Chap5\_DataTransfer\\
    \tab{1} )\\
    \tab{1} Chap9\_Derogations\\
    )
  \end{minipage}
\end{center}

\noindent
Informally, the above expression says that either the requirements of
GDPR Chapters 1--5 are satisfied, or some of the derogations provided by
GDPR Chapter 9 should apply. 
%
In turn, each of the above terms is
equivalent to a compound OWL2 class that captures more details from
the regulation.  Here we illustrate part of the formalization of
GDPR Chapter~2 for an example. \texttt{Chap2\_LawfulProcessing} is
equivalent to the following expression:

\begin{center}
  \begin{minipage}{30em}
    \small\tt
    ObjectUnionOf(\\
    \tab{1} Art6\_LawfulProcessing \\
    \tab{1} Art9\_SensitiveData \\
    \tab{1} Art10\_CriminalData \\
    )
  \end{minipage}
\end{center}

\noindent
The above three conditions apply, respectively, to non-sensitive personal data, sensitive data, and criminal data.  At least one of the three conditions should be satisfied.
In turn, \texttt{Art6\_LawfulProcessing} is defined as:

\begin{center}
  \begin{minipage}{30em}
    \small \tt
 ObjectUnionOf(\\
 \tab{1} ObjectSomeValuesFrom(spl:hasData \\
 \tab{2} SensitiveData\_as\_per\_Art9\\
 \tab{1} )\\
 \tab{1} ObjectSomeValuesFrom(spl:hasData\\
 \tab{2} CriminalConvictionData\_as\_per\_Art10\\
 \tab{1} )\\
 \tab{1} Art6\_1\_LegalBasis\\
 \tab{1} Art6\_4\_CompatiblePurpose\\
 \tab{1} )\\ 
 )
   \end{minipage}
\end{center}

\noindent
Roughly speaking, the above union represents an implication in disjunctive normal form, and should be read like this: if the data involved in the processing is neither sensitive nor criminal conviction data, then either the
fundamental legal bases of Art.~6(1) apply, or the processing is
compatible with the original purpose for collecting the data as per
Art.~6(4).  In order to capture this meaning, class \texttt{Art6\_1}
is defined as:

\begin{center}
  \begin{minipage}{30em}
    \small\tt
 ObjectSomeValuesFrom(hasLegalBasis\\
 \tab{1} ObjectUnionOf(\\
 \tab{2} Art6\_1\_a\_Consent\\ 
 \tab{2}  Art6\_1\_b\_Contract\\
 \tab{2} Art6\_1\_c\_LegalObligation\\ 
 \tab{2} Art6\_1\_d\_VitalInterest\\ 
 \tab{2}  Art6\_1\_e\_PublicInterest\\
 \tab{2}  Art6\_1\_f\_LegitimateInterest\\ 
 \tab{1} )\\
  )\\
     \end{minipage}
\end{center}
  
  \noindent
Roughly speaking, this definition means that a business policy satisfies the requirements of Art.~6(1) if it contains a clause
\begin{center}
\small\tt ObjectSomeValueFrom( hasLegalBasis $X$ )  
\end{center}
where $X$ is some of the above classes corresponding to
points \emph{a--f} of Art.~6(1). In practice, this means that a human
expert has to pick an appropriate legal basis for each business
policy.  Similarly, the formalization of Article~9 applies to sensitive data categories only, and requires a legal basis from a different list.  So the term \texttt{SensitiveData\_as\_per\_Art9} is equivalent to:

 \begin{center}
  \begin{minipage}{30em}
    \small\tt
 ObjectUnionOf(\\
 \tab{1} ObjectSomeValuesFrom(spl:hasData \\
 \tab{2} ObjectComplementOf(SensitiveData\_as\_per\_Art9)\\
 \tab{1} )\\ 
 \tab{1} ObjectSomeValuesFrom(hasLegalBasis\\
 \tab{2} ObjectUnionOf(\\
 \tab{3} Art9\_2\_a\_Consent\\ 
 \tab{3} Art9\_2\_b\_EmploymentAndSocialSecurity\\
 \tab{3} Art9\_2\_c\_VitalInterest\\
 \tab{3} Art9\_2\_d\_LegitimateActivitiesOfAssociations\\ 
 \tab{3} Art9\_2\_e\_PublicData\\
 \tab{3} Art9\_2\_f\_Juducial\\ 
 \tab{3} Art9\_2\_g\_PublicInteres\\
 \tab{3} Art9\_2\_h\_PreventiveOrOccupationalMedicine\\
 \tab{3} Art9\_2\_i\_PublicHealth\\
 \tab{3} Art9\_2\_j\_ArchivingResearchStatistics\\ 
 \tab{2} )\\
 \tab{1}  )\\
 )
   \end{minipage}
\end{center}

\noindent
The rest of the regulation is formalized with a similar
approach.

\subsection{Compliance Checking}

Let us now make an example of compliance checking of a business policy w.r.t.\ the above axiomatization. Consider the following business policy:

\begin{center}
  \begin{minipage}{30em}
    \small\tt
 ObjectIntersectionOf(\\
 \tab{1} ObjectSomeValuesFrom( hasData Religion )\\
 \tab{1} ObjectSomeValuesFrom( hasProcessing Collect )\\
 \tab{1} ObjectSomeValuesFrom( hasPurpose \\
 \tab{2} PersonalisedBenefits )\\
 \tab{1} ObjectSomeValuesFrom( hasStorage \\
 \tab{2} ObjectSomeValuesFrom( hasLocation EU ))\\
 \tab{1} ObjectSomeValuesFrom( hasRecipient\\
 \tab{2} DataProcessor )\\
 \tab{1} ObjectSomeValuesFrom( hasDuty\\
 \tab{2} Art12-22\_SubjectRights )\\
 \tab{1} ObjectSomeValuesFrom( hasDuty\\
 \tab{2} Art32-37\_Obligations )\\
 \tab{1} ObjectSomeValuesFrom( hasLegalBasis\\
 \tab{2} Art6\_1\_a\_Consent )\\
  )
     \end{minipage}
\end{center}

\noindent
This policy is not a subclass of the formalized GDPR (hence it
does not pass the compliance check) because \texttt{Religion} is
classified as sensitive data (it is a subclass
of \texttt{SensitiveData\_as\_per\_Art9}).  Then the business policy
is not a subclass of \texttt{Art9\_SensitiveData}, because the
legal basis is not among the required list.  Moreover, the business
policy is not covered by the derogations provided by GDPR Chapter~9
(details are omitted here).  As a consequence, the business policy does not satisfy the conditions specified by \texttt{Chap2\_LawfulProcessing}.  Note that this kind of compliance checking is able to verify the coherency of the different parts of a business policy. 

If \texttt{Religion} was replaced by any non-sensitive data category
such as \texttt{Location}, then the policy would be compliant because
it would be a subclass of \texttt{Art6\_LawfulProcessing}. This
satisfies the condition called \texttt{Chap2\_LawfulProcessing}.
The \texttt{hasDuty} attributes of the business policy suffice to
satisfy \texttt{Chap3\_RightsOfDataSubjects} and \texttt{Chap4
ControllerAndProcessorObligations}. \texttt{Chap5\_DataTransfer} would also be satisfied since the
processing does not involve any transfers outside the EU.


\section{Our Automated Compliance Checking Algorithm}
\label{sec:compliance}
Business policies (that describe the processing of each of the
controller's processes) are not only needed to fulfill the requirements
of Article~30.  They can also be used to check whether a running
process complies with the available consent, as a sort of access
control system.  Several implementation strategies are possible,
depending on the controller's system architecture; to fix ideas, the
reader may consider the following generic approach: Each of the
controller's processes is labeled with a corresponding business
policy that describes it, and before processing a piece of data, the business policy is
compared with the data subject's consent to check whether the
operation is permitted.

In general, such compliance checks occur frequently enough
to call for a scalable implementation.
Consider, for example, a telecom provider that collects location
information to offer location-based services. Locations cannot be
stored without a legal basis, such as law requirements or consent --
not even temporarily, while a batch process selects the parts that can
be legally kept.  So compliance checking needs to be executed on the
fly.  In order to estimate the amount of compliance checks involved,
consider that the events produced by the provider's base stations are
approximately 15000 per second; the probing records of wi-fi networks
are about 850 millions per day.

In order to meet such performance requirements, SPECIAL has developed
ad-hoc reasoning algorithms for \PL \cite{DBLP:conf/ijcai/Bonatti18},
that leverage \PL's simplicity to achieve unprecedented reasoning
speed.
Compliance checking is split into two phases: first, business policies
are normalized and closed under the axioms contained in the
vocabularies; in the second phase, business policies are compared with
consent policies with a \emph{structural subsumption} algorithm.
We have just completed the evaluation of a sequential Java
implementation of those algorithms, called PLR.  We chose Java to
facilitate the comparison with other engines, by exploiting the
standard OWL APIs, and we refrained to apply parallelization
techniques in order to assess the properties of the basic
algorithms. Before discussing more performant implementation options,
we report the results for PLR.

PLR can pre-compute the first phase, since the business policies are
known in advance and are typically persistent. So the runtime cost is
reduced to structural subsumption.  In this way, on the test cases
derived from SPECIAL's use cases
(cf.\ Table~\ref{pilot-test-cases}), the performance we achieve,
respectively, is $150\mu$sec and $190\mu$sec per compliance check,
using the following system:
\begin{center}
  \small
  \begin{tabular}{ll}
    \hline
    processor: & Intel Xeon Silver 4110 \\
    cores: & 8\\
    cache: & 11M\\
    RAM: & 198 GB\\
    \hline
    OS: & Ubuntu 18.4\\
    \hline
    JVM: &  1.8.0\_181\\
    heap: & 32 GB (actually used: less than 700 MB).\\
    \hline
  \end{tabular}
\end{center}
This means that PLR alone can execute about 6000 compliance checks per
second and more than 518 million checks per day, that is, 60\% of wi-fi probing events and 40\% of
base station events.

\begin{table}
   \begin{center}
     \small
     \begin{tabular}{lrr}
       \hline
      & \multicolumn{1}{c}{\normalsize Pilot~1} & \multicolumn{1}{c}{\normalsize Pilot~2}
       \\
       \hline
       \hline
       \it Ontology & &
       \\
       \hline
       inclusions & 186 & 186
       \\
       disjoint class axioms & 11 & 11
       \\
       property range axioms & 10 & 10
       \\
       functional properties & 8 & 8
       \\
       classification hierarchy height & 4 & 4
       \\
       \hline
       \it Business policies & &
       \\
       \hline
       \# generated policies & 120 & 100
       \\
       avg.\ simple pol.\ per full pol. & 2.71 & 2.39
       \\
       \hline
       \it Consent policies & &
       \\
       \hline
       \# generated policies & 12,000 & 10,000
       \\
       avg.\ simple pol.\ per full pol. & 3.77 & 3.42
       \\
       \hline
       \it Test cases & &
       \\
       \hline
       \# compliance checks & 12,000 & 10,000
       \\
       \hline
     \end{tabular}
   \end{center}
   \caption{Test cases derived from SPECIAL's pilots}
   \label{pilot-test-cases}
\end{table}

In order to raise performance up to the required levels, one
can re-engineer PLR using a language more performant than Java, and/or
parallelize processing by means of big data architectures.
Compliance
checking is particularly well suited to parallelization, since each
test is independent from the others and no synchronization is
required. Additionally, the investigation of parallelization within
PLR's algorithms is under investigation.


\section{Related Work}
\label{sec:related-work}

From a GDPR compliance perspective, there exist several compliance tools (cf. \cite{ICO2017,MicrosoftTrustCenter2017,Nymity,agarwal2018legislative}) that enable companies to assess the compliance of applications and business processes via predefined questionnaires. 
%

%
%

There is also a large body of work on legal knowledge representation (cf.\cite{bartolini2015using,pandit2018gdprtext}) and reasoning (cf. \cite{palmirani2011legalruleml, athan2013oasis, governatori2016semantic, lam2019enabling}).
From a representation perspective, \citet{bartolini2015using} and \citet{pandit2018gdprtext} propose ontologies that can be used to model data protection requirements.
While, \citet{palmirani2011legalruleml} and \citet{athan2013oasis} demonstrate how LegalRuleML can be used to specify legal norms.
The work by \citet{lam2019enabling} and \citet{governatori2016semantic} also builds upon LegalRuleML, however the focus is more on ensuring business process compliance.

Both rule languages and OWL2 have already been used as policy languages; a non-exhaustive list is \cite{DBLP:journals/csec/WooL93,DBLP:journals/tods/JajodiaSSS01,DBLP:conf/policy/UszokBJSHBBJKL03,rei,DBLP:journals/tkde/BonattiCOS10}. As noted in \cite{DBLP:conf/datalog/Bonatti10}, the advantage of OWL2 -- hence description logics -- is that all the main policy-reasoning tasks are decidable (and tractable if policies can be expressed with OWL2 profiles), while compliance checking is undecidable  in rule languages, or at least intractable -- in the absence of recursion -- because it can be reduced to datalog query containment. So an OWL2-based policy language is a natural choice in a project like SPECIAL, where policy comparison is the predominant task.
Among the aforementioned languages, both Rei and Protune \cite{rei,DBLP:journals/tkde/BonattiCOS10} support logic program rules, which make them unsuitable to SPECIAL's purposes. KAoS \cite{DBLP:conf/policy/UszokBJSHBBJKL03} is based on a description logic that, in general, is not tractable, and supports role-value maps -- a construct that easily makes reasoning undecidable (see \cite{DBLP:conf/dlog/2003handbook}, Chap.~5). The papers on KAoS do not discuss how to address this issue.

P3P's privacy policies -- that are encoded in XML -- and simple \PL
policies have a similar structure: the tag \easytt{STATEMENT} contains
tags \easytt{PURPOSE}, \easytt{RECIPIENT}, \easytt{RETENTION}, and
\easytt{DATA\mbox{-}GROUP}, that correspond to the analogous
properties of SPECIAL's usage policies.  Only the information on the
location of data is missing. The tag \easytt{STATEMENT} is included in
a larger context that adds information about the controller (tag
\easytt{ENTITY}) and about the space of web resources covered by the
policy (through so-called \emph{policy reference files}). Such
additional pieces of information can be directly encoded with simple
\PL concepts.

There exist several well-engineered reasoners for OWL2 and its
profiles. Hermit \cite{DBLP:journals/jar/GlimmHMSW14} is a general
reasoner for OWL2. Over the test cases inspired by SPECIAL's use
cases, it takes 3.67~\!ms and 3.96~\!ms per compliance check,
respectively, that is, over 20 times longer than PLR.  ELK
\cite{DBLP:journals/jar/KazakovKS14} is a specialized polynomial-time
reasoner for the OWL2-EL profile. It does not support functional
roles, nor the interval constraints used to model storage duration,
therefore it cannot be used to reason on the \PL profile.  Konclude
\cite{DBLP:journals/ws/SteigmillerLG14} is a highly optimized reasoner
with ``pay-as-you-go'' strategies (i.e.\ it becomes more efficient on
less complex profiles of OWL2).  Konclude is designed for
classification, and is currently not optimized for subsumption tests
(i.e.\ the reasoning task underlying compliance checks). Consequently,
it turns out to be slower than Hermit on our test cases.


\section{Conclusion and Future Work}
\label{sec:conclusions}
The overarching goal of the SPECIAL project is to develop tools and technologies that enable data controllers and processors to comply with personal data processing obligations specified in the GDPR.
In this paper, we presented the SPECIAL policy language and discussed how it can be used to encode consent, business policies, and regulatory obligations. In addition we described the SPECIAL approaches to GDPR compliance checking and presented the results of our initial performance evaluation.

Ongoing/future work includes: the optimisation of the existing compliance checking algorithm to cater for automated compliance checking for a broader set of legislative requirements; and the development of an algebra that can be used to combine multiple policies, for instance where there is a need to aggregate data from multiple data sources.


\section*{Acknowledgment}

This research is funded by the European Union's Horizon 2020 research
and innovation programme under grant agreement N.~731601.
The authors are grateful to all of SPECIAL's partners; without their
contribution this project and its results would not have been
possible.



\bibliographystyle{abbrvnat}
\bibliography{IEEEabrv,D1.1,biblio,services}

\end{document}